\documentclass[journal]{IEEEtran}
\usepackage{graphicx}
\usepackage{amssymb}
\usepackage[cmex10]{amsmath}
\usepackage{cite}
\usepackage{graphicx}
\usepackage[normalem]{ulem}
\usepackage{epstopdf}

\begin{document}
%
\title{Photon Counting OTDR :\\ Advantages and Limitations}

\author{Patrick~Eraerds, Matthieu Legr\'{e}, Jun Zhang, Hugo Zbinden, Nicolas Gisin
\thanks{P. Eraerds, J. Zhang, H. Zbinden, N. Gisin are with the Group of Applied Physics, University of Geneva, 1211 Geneva 4 Switzerland, e-mail: Patrick.Eraerds@unige.ch}
\thanks{M. Legr\'{e} is with idQuantique SA, 1227 Carouge/Geneva, Switzerland}
\thanks{Manuscript received April x, 200y; revised January x, 200y. Financial supports from the Swiss Federal Department for Education and Science (OFES), in the framework of the European COST299 project, from the Swiss NCCR "Quantum photonics" are acknowledged.}}

\markboth{Journal of Lightwave Technology ,~Vol.~x, No.~y, January~200z}%
{Shell \MakeLowercase{\textit{et al.}}: Photon Counting OTDR :\\ Advantages and Limitations}

\maketitle
$\copyright$ 20xx IEEE. Personal use of this material is permitted.
However, permission to reprint/republish this material for advertising or
promotional purposes or for creating new collective works for resale or
redistribution to servers or lists, or to reuse any copyrighted component of
this work in other works must be obtained from the IEEE.\\

\begin{abstract}
We give detailed insight into photon counting OTDR ($\nu-$OTDR) operation, ranging from Geiger mode operation of avalanche photodiodes (APD), analysis of different APD bias schemes, to the discussion of OTDR perspectives. Our results demonstrate that an InGaAs/InP APD based $\nu-$OTDR has the potential of outperforming the dynamic range of a conventional state-of-the-art OTDR by 10 dB as well as the 2-point resolution by a factor of 20. Considering the trace acquisition speed of $\nu-$OTDRs, we find that a combination of rapid gating for high photon flux and free running mode for low photon flux is the most efficient solution.  Concerning dead zones, our results are less promising. Without additional measures, e.g. an optical shutter, the photon counting approach is not competitive.
\end{abstract}



\section{Introduction}
\IEEEPARstart{O}{ptical} Time Domain Reflectometry \cite{Barnoski} is a well known technique for fiber link characterization. Most of today's commercially available optical time domain reflectometers (OTDRs) are based on linear photon detectors, such as p-i-n or avalanche photodiodes (APDs). Although single photon detection features unmatched sensitivity, OTDRs based on this technique ($\nu-$OTDR) \cite{Healey} have reached the market only in niches \cite{ComNUOTDR}.\\
Several single photon detection techniques are possible \cite{HamamatsuPMT}-\cite{Friedrich}, but only few of them are suitable for in-field measurements. Geiger-mode operated InGaAs/InP APDs (for telecom wavelengths) \cite{Ribordy}\cite{Stucki}\cite{Rarity} are the most promising candidates, due to their robustness and manageable cooling.\\
In this paper we discuss the advantages and limitations of these devices, when used in an $\nu-$OTDR. We concentrate in particular on the dynamic range, 2-point resolution, measurement time and dead zone. All $\nu-$OTDR measurements are supplemented by measurements using a conventional state-of-the-art OTDR (Exfo, FTB7600). This makes it easier to evaluate the $\nu-$OTDR performance. Our discussion also contains the possible yield of newly emerged gating techniques like \textit{rapid gating} \cite{Shields}\cite{Inoue}\cite{Jun2}.\\
We note that some years ago $\nu$-OTDRs based on silicon APDs, suitable for C-band operation, were demonstrated \cite{Diamanti}\cite{Legre}. Although silicon APDs show superior behavior, concerning afterpulsing and timing jitter, the upconversion of telecom photons to the visible regime demands more expensive optics and more sophisticated alignment. Therefore we believe that they are less suitable when robustness is required.\\
Paper organization : In Sect.II we provide information about Geiger-mode operation of InGaAs/InP APDs and discuss its major impairment, the afterpulsing effect. Sect.III focusses on $\nu-$OTDR operation and performance (dynamic range, 2-point resolution, measurement time, dead zone) and compares it with the performance of a conventional state-of-the-art long haul OTDR (Exfo FTB7600). Sect.IV considers time efficient bias schemes (\textit{rapid gating, free running}) and finally we summarize our results in Sect.V.
\section{Geiger-mode APD}
\subsection{Basic operation}
\label{Basic operation}
In Geiger-mode, the APD is biased beyond its breakdown voltage, typically by a few percent. This provides a sufficiently large gain (order of $10^6$) to detect a single incident photon (with detection efficiency $\eta$). In contrast to a linear mode APD, the output signal is no longer proportional to the number of primary charges. Whenever an avalanche occurs and the current reaches a certain discrimination level, a detection is counted, independent of how many primary charges caused or were created during the avalanche.\\
To reset the APD for the next detection, the avalanche needs to be quenched. This is typically done by lowering the bias voltage, either actively or passively \cite{Cova}.\\
An APD based on InGaAs/InP  is particularly well suited for use with the principal telecom wavelength bands. Although the dark count\footnote{A detection which was not initiated by a signal photon but thermal excitation or tunneling.} rate is higher than in silicon based APDs, high sensitivity can be regained by cooling, typically around $-50^\circ$C (see also Sect.\ref{DetSens}).\\
There are different ways of  applying the overbias ($V_{bias}>V_{bd}$ (breakdown voltage)) to the diode. The most common ones are the \textit{gated mode} and the \textit{free running mode} \cite{Thew}. In \textit{gated mode} the overbias is applied only during a short time $\Delta t_{gate}$ (called gate), in a repetitive manner with frequency $f_{gate}$ (respecting $f_{gate}<\frac{1}{\Delta t_{gate}})$. Typically $\Delta t_{gate}\in [2 \mbox{ns}, 20\mu\mbox{s}]$ and $f_{gate}\in [100\mbox{Hz},10\mbox{MHz}]$. In \textit{free running mode}, the overbias is applied until a photon or noise initiates an avalanche.\\ While the \textit{gated mode} achieves high signal to noise ratios when a synchronized signal is being detected, the \textit{free running mode} is most suited when the photon arrival time is not known (e.g. in OTDR).\\
More recent developments, summarized by the name \textit{rapid gating} \cite{Shields}\cite{Inoue}\cite{Jun2}, apply very short gates ($\approx$ 200 ps) in order to severely limit avalanche evolution and reduce afterpulsing (see Sect.\ref{Afterpulsing}). The technical challenge consists in discriminating the rather small avalanche signal from the capacitive response to overbias of the diode itself. In "classical gating", described in the previous paragraph, one usually waits until the avalanche signal is easy to discriminate. Typical gating frequencies in \textit{rapid gating} are of the order of 1 GHz.\\
In Sect.IV we will discuss pros and cons of these different approaches, in particular concerning their applicability for $\nu-$OTDRs.\\
\subsection{Detection sensitivity}
\label{DetSens}
\begin{figure}[t]
\includegraphics[width=8cm]{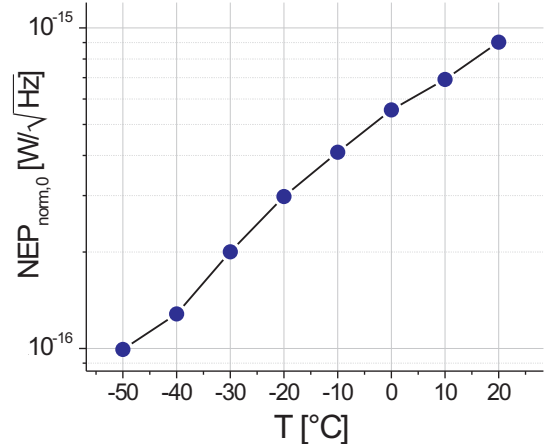} 
\caption{Bandwidth normalized noise equivalent power ($NEP_{norm,0}$, see Eq.\ref{NEP0Geiger}) as function of Geiger-mode APD temperature. The detection efficiency $\eta$ is kept constant at 10\%. At ambient temperatures the noise equivalent power is increased by almost a factor of 10 with respect to the usual operating temperature of $-50^\circ$C.}
\label{NEPTEMP}
\end{figure}

A figure of merit for the sensitivity of a detector is its noise equivalent power ($NEP$). For example, the bandwidth normalized $NEP_{norm}$ of a linear photo detector is given by \cite{Hamamatsu}\cite{Perkin}  \\
\begin{equation}
NEP_{norm} = \frac{\Delta I_{noise}}{S\cdot G}  \quad [\frac{\mbox{W}}{\sqrt{\mbox{Hz}}}]
\end{equation}\\
where $\Delta I_{noise}$ ~[A/$\sqrt{\mbox{Hz}}$] is the standard deviation of the total noise current (thermal-, dark-, signal shot- and in case of gain also gain noise), normalized with respect to the bandwidth of the detector, $S$~ [A/W] is the detector photosensitivity and $G$ is the gain of the diode (p-i-n diode : $G=1$, linear APD (typically) : $G=10-100$).\\
APDs are superior to p-i-n diodes in the circuit noise limited regime\footnote{Circuit noise results for example from thermal motion of charges in resistors or charge fluctuation in transistors in the receiver amplifier.} \cite{Teich}, but lose their advantage when the gain noise becomes important, i.e. at stronger signal powers. The minimal detectable power $NEP_{norm,0}~$[W/$\sqrt{\mbox{Hz}}$] is obtained by setting the signal power and thus the signal shot-noise equal to zero. $NEP_{norm,0}$ can usually be found in the data sheet of the diode, typically $10^{-15}- 10^{-13} ~$[W/$\sqrt{\mbox{Hz}}$] for InGaAs APDs at $25^\circ$C.\\
A similar expression can be derived for Geiger-mode APDs (see App.\ref{APPC}, Eq.\ref{NEPnorm}):\\
\begin{equation}
NEP_{norm}=\frac{h\nu}{\eta}\cdot\sqrt{2\cdot\hat{p}_{noise}}\quad\quad [\frac{\mbox{W}}{\sqrt{\mbox{Hz}}}]
\label{NEPGeiger}
\end{equation}\\
where $\eta$ is the detection efficiency and $\hat{p}_{noise}$ is the noise detection probability per gate (including signal and dark count shot noise), normalized  with respect to the gate width $\Delta t_{gate}$ in seconds. Again, setting the input optical power equal to zero, we infer the minimal detectable power (App.\ref{APPC}, Eq.\ref{NEPnorm0})
\begin{equation}
NEP_{norm,0}=\frac{h\nu}{\eta}\cdot\sqrt{2\cdot\hat{p}_{dc}}\quad\quad [\frac{\mbox{W}}{\sqrt{\mbox{Hz}}}]
\label{NEP0Geiger}
\end{equation}\\
where $\hat{p}_{dc}$ denotes the dark count probability per gate, normalized with respect to the gate width $\Delta t_{gate}$ in seconds. Inserting the parameters of the Geiger mode APD used in our experiments ($\hat{p}_{dc}=2000~\mbox{s}^{-1}$, $\eta=10$\%, $T=-50^\circ$C, Sect.\ref{DynamicR}), we estimate $NEP_{norm,0}\approx 10^{-16}$~[W/$\sqrt{\mbox{Hz}}$].\\
In Fig.\ref{NEPTEMP} we see the evolution of $NEP_{norm,0}$ as function of temperature. We observe that when approaching ambient temperatures, we almost reach the regime of the best linear mode diodes.
Conversely, one might be tempted to cool linear diodes to $-50^\circ$C to reach the $NEP$ of Geiger mode APDs. Even if this might be in general achievable, one should not forget, that the output signal still needs to be amplified. Even at ambient temperatures the small pulse amplifier noise usually constitutes the dominating noise source leading to much higher effective NEPs.\\
By analyzing the performance of a conventional OTDR in Sect.III, we will gain more insight into the sensitivity limits of linear mode APD detection systems.
\subsection{Afterpulsing}
\label{Afterpulsing}
\begin{figure}[h]
\includegraphics[width=\linewidth]{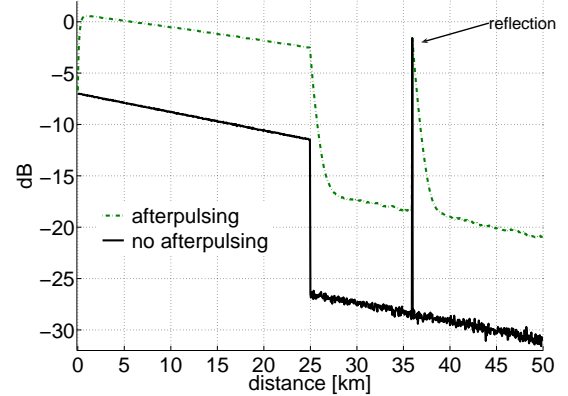} 
\caption{Illustration of the afterpulsing effect on $\nu-$OTDR trace. Most severe are the dead zones after large loss events (at 25 km) and reflections (at 35 km). More subtle is the change of the slope of the trace which is usually smaller than what is measured when afterpulsing can be neglected.}
\label{APtrace}
\end{figure}

One of the major impairments of InGaAs/InP APDs is the afterpulsing effect. Imperfections and impurities in the semiconductor material are responsible for intermediate energy levels (also called trap levels), located between the valence band and the conduction band. During an avalanche, these levels get overpopulated with respect to the thermal equilibrium population. If the APD gets reactivated right after the quenching of an avalanche, the probability of thermal excitation or tunneling of one of these charges into the conduction band and the subsequent initiation of an afterpulse avalanche, is high. Although fundamentally the improvement of semiconductor purity and thus the reduction of the number of trap levels is preferable, different mitigation measures can be carried out :\\
\textit{a) dead time }: A purely passive measure is the introduction of a dead time. The trap population decreases exponentially with time, due to thermal diffusion. Finally the thermal equilibrium configuration is restored. The impact of afterpulsing can therefore be mitigated by maintaining the bias voltage below breakdown, i.e. the application of a dead time $\tau$, after a detection takes place. Dead times severely limit the maximum achievable detection rate.\\
\textit{b) heating }: An increased temperature accelerates the diffusion of trapped charges. However, at the same time charge excitation from the valence into the conduction band increases, leading to globally increased noise, which eventually reduces the detector sensitivity. Thus one cannot achieve low afterpulsing and high sensitivity at the same time. It is necessary to find a trade-off depending on the particular application.\\
\textit{c) quenching technique} : As soon as the avalanche has gained enough strength such that the current pulse can be detected, it needs to be quenched. The quenching speed is crucial to limiting the number of secondary charges which can populate trap levels. Here, fully integrated active quenching circuits yield much better results than non-integrated ones \cite{Jun}. Another approach is \textit{rapid gating} (Sect.\ref{Basic operation}). Avalanche evolution is terminated by short gate durations (200 ps) and the number of secondary charges is kept low.\\
\begin{figure}
\includegraphics[width=\linewidth]{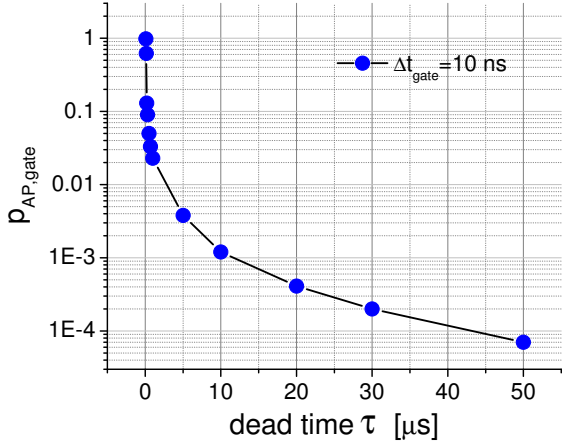} 
\caption{Afterpulse probability as function of dead time $\tau$. The detection efficiency $\eta$ is equal to 10\% and the temperature T=$-50^\circ C$. An active quenching application specific integrated circuit (ASIC) \cite{Jun} was used.}
\label{APCurves}
\end{figure}\\
In Fig.\ref{APCurves} we plot an example of afterpulse probability as function of dead time $\tau$, using a fully integrated ASIC based active quenching circuit \cite{Jun}. Whenever a detection takes place, we activate a second gate of width $\Delta t_{gate}= 10~ns$ with a temporal delay of $\tau$. In this second gate, no incident photons are present. If there is a detection it is either a dark count or an afterpulse. Since for large $\tau$ only the actual dark counts remain, we can subtract it from the total count rate and obtain the pure afterpulsing probability ($\rightarrow$ Fig.\ref{APCurves}). During larger gates, the afterpulse probability sums up and afterpulsing increases. One can easily calculate the afterpulse probability of a gate of width  $\Delta t_{gate}$ ($\leq 10 \mu s$) by
\begin{equation}
	p_{AP,\Delta t_{gate}}(\tau)=1-(1-p_{AP,10 ns}(\tau))^m \\
\label{InferingAP}
\end{equation}
where $p_{AP,10 ns}(\tau)$ is the afterpulse probability in a gate of 10 ns width and $m=\frac{\Delta t_{gate}}{10ns}$.\\
We note that if no afterpulse occurs in the first activated gate after a detection, it can also happen in any succeeding gate. However, the probability decreases due to trap charge diffusion. To get the total afterpulse probability, or rather the signal to afterpulse ratio, one needs to account for this summing effect as well. The lower the signal detection rate, the more summing-up takes place. Thus a higher signal detection rate improves the signal to afterpulse ratio.\\
In Fig.\ref{APtrace} we illustrate the impact afterpulsing can have in a $\nu-$OTDR measurement.
Firstly and most importantly we must consider dead zones (see Sect.\ref{DEADZ} for definition, not to be confused with dead time). Whenever an important loss (at 25 km) or a reflection (at 36 km) occurs, it is followed by a tail which prevents the detection of the Rayleigh backscatter directly behind it. Secondly, the backscatter trace is shifted to higher values, since more detections than in the pure signal case occur (pile-up effect). Thirdly, the slope of the trace is flatter than it should be.

How much afterpulsing can be tolerated, generally depends on the particular measurement. For instance in a coarse measurement on a long span of fiber, where only peak positions or large loss events are of interest, one can tolerate a fairly high afterpulsing contribution. On the other hand, in the case of short links, where high precision for fiber attenuation measurement and dead zone minimization is desired, afterpulsing must be kept to a few percent or even lower, depending on required precision of the measurement.\\
Numerical afterpulse correction methods were also analyzed \cite{Wegmuller}, but it was found that for high precision measurements the algorithm lacks robustness due to possible variations in the afterpulse probability. It should therefore be used only when the requirements on precision are not stringent.\\
\section{Photon counting vs. conventional OTDR}
\begin{figure}[h]
\centering
\includegraphics[width=\linewidth]{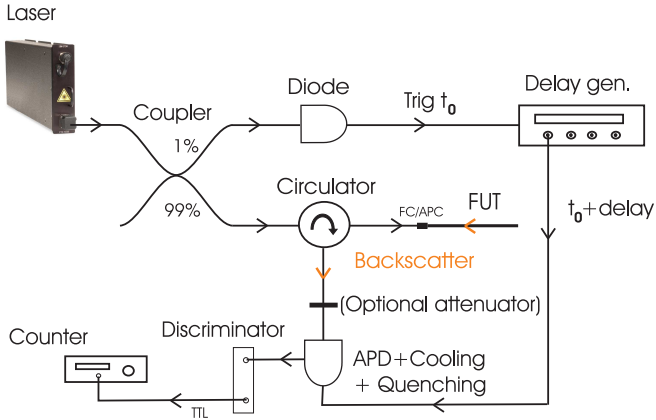}
\caption{Basic $\nu$-OTDR setup. The laser (we use the laser of the FTB-7600, $P_{peak}=400$ mW) emits pulses with a frequency $f_{pulse}$ adapted to the length of the fiber under test $L_{fiber}$ ($\rightarrow f_{pulse}= \frac{c}{2\cdot L_{fiber}}$). The signal is split at a 99/1-coupler. The 99\% part is launched into the fiber under test (FUT) via a circulator. Backscattered light from the fiber exits the lower port of the circulator and illuminates the InGaAs/InP-APD. The 1\% part is used to measure the time of departure $t_0$ of the laser pulse (for synchronization reasons) using a conventional photodiode (Newport, 1GHz). The output signal is sent to a delay generator. A delayed signal at $t_0+t_{delay}$ is sent to the APD to apply a detection gate of length $\Delta t_{gate}$ and the backscattered intensity corresponding to the applied delay is measured. The APD reverse bias is equal 48.7 V, yielding a detection efficiency $\eta = 10$\% at $-57^\circ$C (minimal value). We measure a dark count probability per gate equal to $\hat{p}_{dc}= 2000~\mbox{s}^{-1}$ (normalized with respect to the gate width,  Eq.\ref{NEP0Geiger}). This corresponds to a dark count probability of $2\cdot10^{-5} $ for a gate width of 10 ns and $2\cdot10^{-2} $ for a 10 $\mu$s gate.}
\label{setup}
\end{figure}

Although this section is mainly concerned with the $\nu$-OTDR technique, we also perform measurements using a state-of-the-art conventional\footnote{Based on linear-mode APD detection.} OTDR (FTB-7600, EXFO), a product especially designed for long-haul applications (up to 50 dB dynamic range). This makes it easier for us to highlight the advantages and drawbacks of the photon counting approach.
The experimental setup and a detailed explanation is given in Fig.\ref{setup}.
\\
For a fixed delay, a number of laser pulses $N_{pulse}$ (repetition frequency $f_{pulse}$) are sent and $N_{gate} (=N_{pulse}$) gates are activated. A counter records the number of detections. The incident signal power can be inferred from the ratio of detections to activated gates (for details, see App.\ref{APPA}).\\
To get information on the backscatter of the entire fiber, the delay needs to be scanned, repeating the procedure explained before for each single delay position. The sampling resolution, i.e. the delay step $\Delta t_{delay} ~(t_{delay}=i\cdot\Delta t_{delay}, i=1,2,3...)$, needs to be adapted to the requirements of the particular measurement (e.g. zooming or coarse full trace measurement). The detection bandwidth is given by $B=\frac{1}{2\Delta t_{gate}}$.\\
We note that this is only the most basic version of a photon counting OTDR. One of the advantages of this system is that due to the low gating frequency, we can totally exclude afterpulse effects (dead time 2 ms). Therefore we can determine the unadulterated dynamic range and 2-point resolution. Nevertheless, data acquisition is very time consuming. For example the measurement of the entire 200 km fiber, discussed in the next section (Fig.\ref{Comparison1mus}), took about 6 hours. In Sect.IV we will see, how it can be performed more efficiently.
\subsection{Dynamic range}
\label{DynamicR}
\begin{figure}[h]
\includegraphics[width=\linewidth]{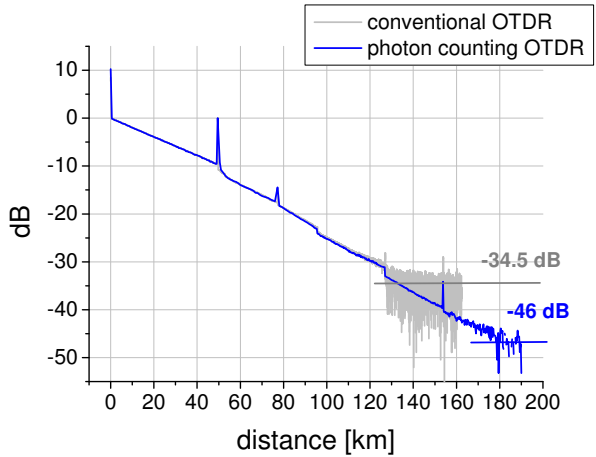} 
\caption{OTDR traces of 200 km fiber link using a laser pulse width of $1\mu s$. The light gray curve represents the result of the conventional OTDR (Exfo FTB-7600) after 3 minutes of measurement in standard configuration. In this configuration it uses a detection band width of 4 MHz at the end of the measurement range. The $\nu-$OTDR result is represented by the blue curve. The measurement bandwidth is 500 kHz using the same number of samplings for each point as the conventional device. The bandwidth adapted results for different pulse widths can be found in Fig.\ref{DynRanges}.}
\label{Comparison1mus}
\end{figure}

\begin{figure}
\centering
\includegraphics[width=8cm]{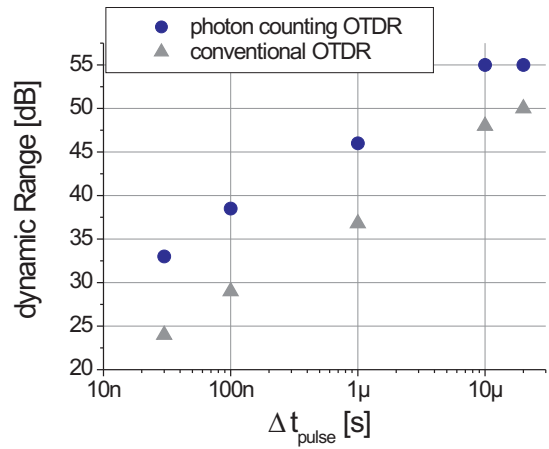} 
\caption{Dynamic ranges of FTB-7600 and $\nu-$OTDR for different laser pulse widths. The length of the used fiber was 200 km. The detection bandwidths $B$ were adapted in each case, it holds that $B=\frac{1}{2\Delta t_{pulse}}$.}
\label{DynRanges}
\end{figure}

To measure the dynamic range of both devices for different laser pulse widths, we take a 200 km fiber, composed of a 50 km spool and an installed fiber link of 150 km (Swisscom, Geneva-Neuchatel), which itself consists of several fibers. The length of the fiber allows a maximal laser pulse repetition rate of $f_{laser}=\frac{c}{2\cdot L_{fiber}}=500$ Hz, where $c$ is the speed of light in standard optical fiber.\\
We start measuring the trace with the FTB-7600 for 3 minutes\footnote{We choose 3 min because this is the time specified in the definition of OTDR dynamic range for conventional OTDRs \cite{Derickson}.} with a laser pulse width of 1~$\mu$s. The device acquires $180~s\cdot500~\mbox{Hz}=9\cdot10^4$ different traces. The final output trace is the numerical average of these single traces (Fig.\ref{Comparison1mus}, light grey curve). For a fair comparison the detection bandwidth should be equal for the two devices. For the conventional OTDR it is automatically chosen by the device and not available to us. We infer its value by looking at the noise period at the end of the measurement range. For a pulse width of $1\mu s$ we obtain 4 MHz. Under these conditions the dynamic range is found to be 34.5 dB.\\
We then perform the $\nu$-OTDR measurement, ensuring that we do not saturate the detector with the backscatter from the beginning of the fiber. Therefore we insert an additional attenuator in front of the APD to reach the unsaturated regime. We adjust the attenuation to yield a detection rate of about 90\% of the gate rate for the first delay position. At each delay position we count the number of detections within 3 minutes, which yields the same statistics per sampling point as in the previous case ($180~s\cdot500~\mbox{Hz}=9\cdot10^4$ samplings). We choose $\Delta t_{delay}$ equal to 3 $\mu$s (=300 m sampling point separation in fiber) and $\Delta t_{gate}=\Delta t_{pulse}$. With increasing delay the backscatter power and thus the detection rate decreases. When we start to approach the noise level of the detector, we pause the measurement and remove a part of the attenuation (to regain 90\% detection rate), reduce the delay for a few kilometers (to get an overlap with the previous part) and resume the measurement. In this way we obtain several single traces of adjacent parts of the fiber. In the following we will refer to this as partial trace measurement. By means of the overlaps, the entire trace can be reconstructed. Each partial trace measurement contributes approximately 20 dB to the overall $\nu$-OTDR dynamic range. For example, to cover 50 dB of fiber loss, we need to perform three partial trace measurements\footnote{The first measurement covers 0-20 dB, the second 15-35 dB and the third 30-50 dB respecting the necessary overlap between different partial trace measurements.}. \\
The result of the $\nu$-OTDR measurements is also shown in Fig.\ref{Comparison1mus} (blue curve). It is important to note that we adapt the gate width to the laser pulse width to obtain the minimal $NEP_0$ (App.\ref{APPC}, Eq.\ref{NEP0}) without affecting the 2-point resolution (limited by the laser pulse width). This means that in the case of Fig.\ref{Comparison1mus}, the detection bandwidth of the $\nu-$OTDR is $B =\frac{1}{2\cdot\Delta t_{gate}}= 500$ kHz\footnote{The conventional device uses a higher bandwidth since higher sampling resolution is useful when the position of an event needs to be determined with higher precision.}. To be able to compare the measured results in a representative manner, we average the conventional OTDR trace in order to obtain the same bandwidth as was used in the $\nu-$OTDR measurement. We gain 2.5 dB, yielding a corrected dynamic range of 37 dB. The $\nu-$OTDR advantage is found to be roughly 9 dB in this case.\\
We repeat the measurement for different pulse widths keeping all other parameters unchanged. The final results are shown in Fig.\ref{DynRanges}. The detection bandwidths were adapted as before. It holds that $B=\frac{1}{2\Delta t_{pulse}}$ for both devices.\\
For pulse widths between 30 ns and 1 $\mu$s the dynamic range difference is about 9-10 dB. This is a direct consequence of the smaller $NEP_{norm,0}$ (see Sect.\ref{DetSens}) of the Geiger-mode APD, since bandwidth and integration time per sampling point were equally chosen. This means that the $NEP_{norm,0}$ of the $\nu-$OTDR is roughly a factor 63-100 smaller\footnote{Respecting the functional dependence between NEP and dynamic range given in Eq.\ref{EquDynR}, using $NEP_0\propto NEP_{norm,0}$} than the $NEP_{norm,0}$ of the conventional OTDR (noise dominated by the small pulse amplifier).
However, going to larger laser pulses, we observe increased noise for the $\nu$-OTDR and the advantage gets smaller. We suppose that this happens due to the increased backscatter power from the beginning of the fiber. Although the diode is not active, the charge persistence effect (also sometimes called charge subsistence) can have a non negligible impact on the noise counts in a subsequently activated gate (for more details see also Sect.\ref{DEADZ}).\\
In summary, by adapting sampling statistics and detection bandwidth of both devices we find an advantage of about 9-10 dB in dynamic range for the $\nu$-OTDR. By increasing the laser pulse width we observe increased detector noise and the effective advantage gets smaller. We uncouple the question of measurement time since it is highly dependent on the gating technique used in the $\nu-$OTDR. This discussion is postponed to section \ref{Timeeff}.
\subsection{2-point resolution}
\begin{figure}
\centering
\includegraphics[width=\linewidth]{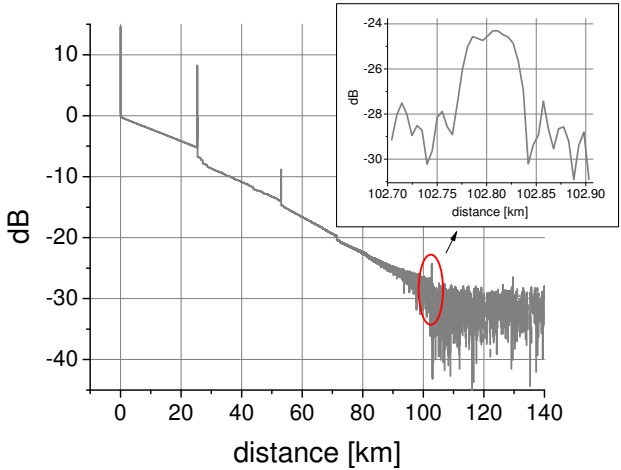} 
\caption{OTDR trace measured using the FTB-7600 with a laser pulse width of 300 ns, acquisition time = 3min. The achieved dynamic range is about 30 dB. At the transition to the noise level (at about 102.8 km), we can see a reflection peak (maybe two, see inset). The FTB-7600 cannot achieve a better resolution of the peak without decreasing the laser pulse width. As the dynamic range would drop (keeping the acquisition time unchanged) and the peak would not be seen anymore.}
\label{LimitTrace}
\end{figure}

\begin{figure}
\centering
\includegraphics[width=\linewidth]{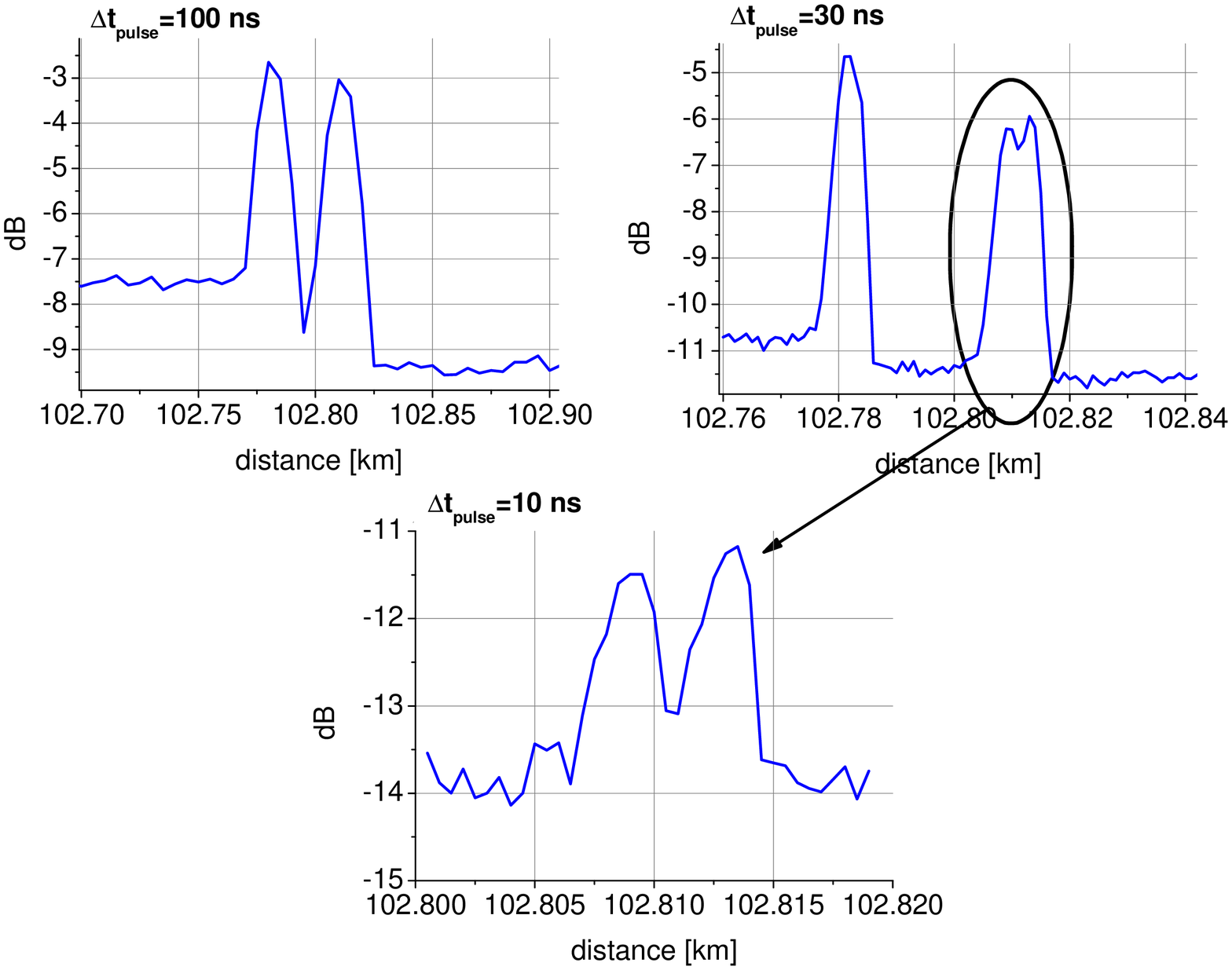} 
\caption{Step by step reduction of laser pulse width in $\nu-$OTDR measurement, when zooming on the reflective event seen by the FTB-7600 (Fig.\ref{LimitTrace}). The lower graph is again a zoom on the two reflective events which were revealed by the second graph (peak 2 and 3). Due to its larger sensitivity the $\nu-$OTDR can afford smaller laser pulses at distances where the FTB-7600 reaches its limits.}
\label{Zoom}
\end{figure}

When considering the 2-point resolution\footnote{By 2-point resolution we mean the minimal distance, necessary between two reflective events, in order to be able to recognize them as distinct peaks on the OTDR trace output (e.g. dip between peaks at least 1 dB lower than the peaks themselves)}, one can divide OTDR operation into two regimes a) the receiver limited and b) the laser peak power limited regime. In case a) the ultimate timing resolution is either given by the amplifier bandwidth or the detector jitter (using fine laser pulses), whereas in case b) the limited laser peak power makes it necessary to use larger pulse widths (larger than the limit given in case a)) in order to reach high dynamic ranges. In the receiver limited regime, the advantages of photon counting were already discussed in \cite{Wegmuller}, yielding a maximal 2-point resolution of 10 cm for the $\nu$-OTDR and 1 m for the conventional device. In long haul OTDR applications, we operate in the laser peak power limited regime.\\
It is easy to see that in this regime the sensitivity advantage ($NEP_{norm,0}$) of photon counting translates directly into an advantage in 2-point resolution.\\
Example : we consider a reflective event at the end of the dynamic range (with a certain pulse width and measurement time) of the FTB-7600, see Fig.\ref{LimitTrace}.\\
The $\nu-$OTDR can achieve the same dynamic range with a much smaller pulse width, see Fig.\ref{DynRanges}. According to App.\ref{APPE}, an advantage of 10 dB in dynamic range\footnote{Bandwidth normalized.} corresponds to an advantage in 2-point resolution by a factor of 20.
In Fig.\ref{Zoom} we see the results of three $\nu-$OTDR measurements focusing on the reflective event at about 102.8 km.  With each reduction in pulse width more and more structure is revealed and we actually find 3 reflections. The ratio of the peak widths agrees well with the calculated one.
\\
In summary, when the OTDRs are operated in the laser peak power limited regime, the sensitivity advantage of the $\nu-$OTDR translates directly into an advantage in 2-point resolution. Its amount is described by Eq.\ref{2pointAdv} in App.\ref{APPE}.
\subsection{Measurement time}
\label{meastime}
We start discussing the measurement time by taking a look at the time necessary to obtain a sufficient signal to noise ratio (SNR) for a specific delay position in the fiber, from which we receive a backscatter power $P_{opt}$ (see App.\ref{APPF}):
\begin{equation}
	t=\frac{1}{f_{pulse}}\cdot\left(\frac{SNR\cdot NEP_{norm}\cdot\sqrt{B}}{P_{opt}}\right)^2
	\label{tmeas}
\end{equation}
where $NEP_{norm}$ (see Eq.\ref{NEPnorm}) is the noise equivalent power normalized with respect to detector bandwidth in [W/$\sqrt{\mbox{Hz}}$], $B$ the detector bandwidth in [Hz] and $f_{pulse}$ the laser pulse repetition rate in [Hz]. This formula applies to both the Geiger and linear mode operation as long as linearity between input and output signal is guaranteed\footnote{Linearity in Geiger mode applies if the signal detection probability per gate $p_{sig,gate}$ depends linearly on the input power $P_{opt}$, see App.\ref{APPA} Eq.\ref{detProb} \& \ref{muPopt}. For sufficiently small $P_{opt}$ it holds that : $p_{sig,gate}=\eta\cdot\frac{P_{opt}\cdot\Delta t_{gate}}{h\nu}$}.\\
While the Geiger mode exhibits linearity only in a relatively restricted domain of $P_{opt}$, the linear mode is able to cover several orders of magnitude of input power. Therefore Eq.\ref{tmeas} is applicable for a much wider range of optical powers and shows the significant advantage of the linear mode when larger powers need to be measured ($\propto P_{opt}^{-2}$).\\
More interesting from the $\nu-$OTDR perspective is the case when Eq.\ref{tmeas} applies as well for Geiger-mode, i.e. for sufficiently small powers (order -100 dBm). We can then easily calculate the ratio of measurement times (assuming $f_{pulse},B$ and $P_{opt}$ to be equal):
\begin{equation}
	\frac{t^{(conv)}}{t^{(pc)}}=\left(\frac{NEP_{norm}^{(conv)}}{NEP_{norm}^{(pc)}}\right)^2
	\label{timeratio}
\end{equation}
where the superscript $pc$ signifies photon counting (i.e. Geiger mode), and $conv$ represents the conventional detection mode (i.e. linear mode).\\
$NEP_{norm}$ can be split into a signal initiated noise contribution $NEP_{norm,sig}$ (e.g. due to signal shot noise) and a contribution from signal independent sources (dark current, dark counts) represented here by $NEP_{norm,0}$ :
\begin{equation}
	NEP_{norm}^2=NEP_{norm,sig}^2+NEP_{norm,0}^2
\end{equation}
For sufficiently small input power, $NEP_{norm}$ becomes $NEP_{norm,0}$. In Sect.\ref{DynamicR} we estimated the ratio of the $NEP_{norm,0}$ of conventional and $\nu-$OTDR to lie within 63 and 100. Thus the ratio given in Eq.\ref{timeratio} approaches a value between 4000 and 10000. This means that we continuously pass from a huge linear mode advantage ($P_{opt}$ large) to a huge Geiger mode advantage ($P_{opt}$ small).\\
We stress at this point that this result applies for the measurement of a specific delay position. OTDR measurements consist of scanning a range of different powers, i.e. the exponentially decreasing backscatter power. In case of the conventional OTDR the achievement of a sufficient SNR for a certain delay position, e.g. far down the fiber, means that all delays closer to the beginning of the fiber have been scanned at least with the same SNR. Thus we already obtain the full trace up to that point. How large the actually scanned interval is in the case of photon counting depends on the gating technique. For example, using the basic approach, explained earlier  (Fig.\ref{setup}), one would scan exactly one position. In Sect.IV we discuss how the NEP advantage of photon counting can be used more efficiently.\\
\subsection{Dead zone}
\label{DEADZ}
Dead zones are parts of a fiber link where the OTDR trace does not display the actual Rayleigh backscatter, but a signal induced by another source. One example that we have already encountered is the afterpulsing effect. If not accounted for it leads to tails after large loss events or reflections (see Fig.\ref{APtrace}). Unfortunately this is not the only origin of dead zones. If we mitigate the afterpulsing effect by using an appropriate dead time, another effect, very similar to afterpulsing gets dominant : charge persistence. Even though the APD is not biased beyond breakdown between adjacent gates, there is still a bias which can weakly multiply primary charges created by photons incident during that time. These weak avalanches might however lead to increased trap population and increased noise avalanche probability in the next gate ($V_{bias}>V_{bdown}$). This effect, although less severe than afterpulsing, becomes visible under the same circumstances, namely after large loss events and reflections.\\
\begin{figure}[b]
\includegraphics[width=9cm]{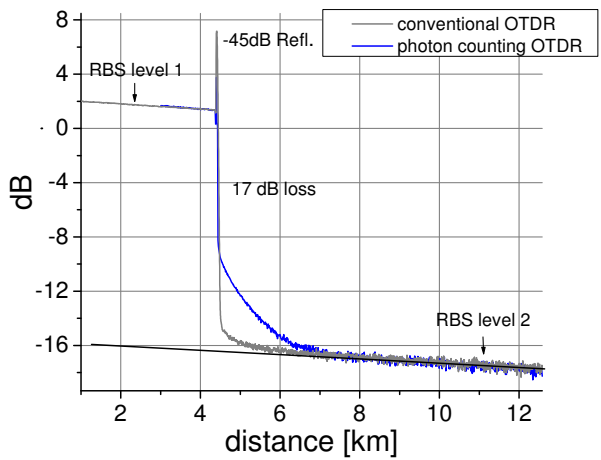}
\caption{Behavior of conventional and photon counting OTDR when subjected to significant change in backscatter power level (here : 17 dB). We introduced a reflection of -45dB in front of the loss, simulating for example a bad connector. The black line represents a fit of the backscatter behind 8 km and can be used as a reference to assess the magnitude of the dead zone.}
\label{deadzone}
\end{figure}
To estimate the impact of this effect on the OTDR output, we perform a measurement on a fiber link containing a large loss event (17 dB), with a  reflection (-45 dB) just before it. This link simulates a typical situation encountered in passive optical networks, where a splitter of high multiplicity induces a significant loss. Weak reflections right in front can be induced by bad connectors.\\
We perform a measurement of this particular fiber link situation, using our $\nu-$OTDR in basic mode, which ensures that no afterpulsing effect is present and the charge persistence effect becomes visible. Our results, including the measurement using the conventional OTDR, are shown in Fig.\ref{deadzone}. We observe the emergence of a tail approximately 10 dB below the loss edge, which decays by approximately 3.5 dB/km. The lower Rayleigh backscatter level gets visible again after about 2 km ($20\mu s$).\\
The result obtained with the conventional OTDR is much better. The emergence of the tail starts on a significantly lower level. Full sensitivity is regained after 1 km.
The results found for the $\nu-$OTDR, confirm the observations made in \cite{Wegmuller} and \cite{Scholder}. One possibility to mitigate dead zones, induced by charge persistence, is the use of an optical shutter as performed in \cite{Scholder}. If the initial backscatter is blocked by the shutter and the gate gets activated, as well as the shutter deactivated right after the loss event, much better results can be obtained.\\
In summary, concerning dead zones, the conventional OTDR shows superior behavior, when no additional measures are taken in the case of the $\nu-$OTDR, e.g. using an optical shutter.
\section{Time efficient bias schemes}
\label{Timeeff}
\begin{figure}[h]
\includegraphics[width=\linewidth]{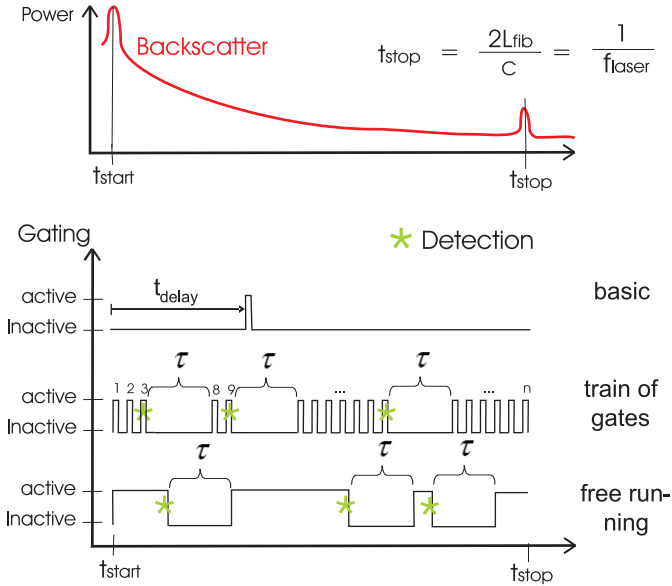}
\caption{Different bias schemes for backscatter measurement : basic, train of gates and free running mode. $\tau$ represents the dead time.}
\label{GSchemes}
\end{figure}

\begin{figure}[t]
\centering
\includegraphics[width=\linewidth]{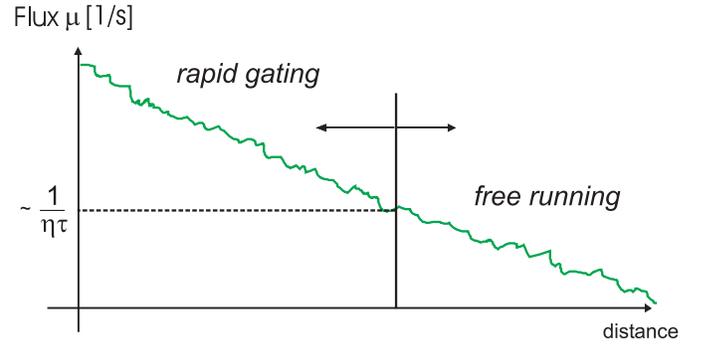}
\caption{According to Eq.\ref{fdet} we can determine the flux regimes where each of the techniques at disposal deliver their best performance. In the low flux regime $\mu<\frac{1}{\eta\cdot\tau}$ the detection duty cycle is most important, therefore the \textit{free running mode} is the optimal solution. As soon as the dead time becomes the limiting factor for the detection rate, \textit{rapid gating} can take advantage of its significantly lower afterpulsing, resulting in much smaller required dead times.}
\label{Regions}
\end{figure}

The way we implemented the $\nu$-OTDR in Sect.III is one of the simplest and trace acquisition is time consuming. It is well suited to study general characteristics, but not for other applications. Its apparent drawback is the wasting of backscattered signal, due to the fact that $f_{gate}=f_{pulse}$, i.e. only one gate per laser pulse is activated.\\
A more efficient approach is the \textit{train of gates} scheme \cite{Wegmuller}. Unlike to the basic mode, the gating frequency is higher than the laser pulse repetition rate and more than one gate gets activated per laser pulse, see Fig.\ref{GSchemes}.
In the ideal case we could choose a gating frequency $f_{gate}$ in the way that the designated sampling resolution is obtained. Unfortunately we have to account for the afterpulsing effect and thus need to apply a dead time $\tau$ (whose length depends on the tolerable afterpulsing)\footnote{Like depicted in Fig.\ref{GSchemes}, we can number each gate in the train of gates, $1...n$. In the ideal case (no dead time) each of these gates gets activated for each laser pulse sent into the fiber. In reality, when a dead time needs to be applied, it is not certain that gate i gets activated. It is possible that it falls into the dead time application range (gate 4-7 in Fig.\ref{GSchemes}).}. In App.\ref{APPG} we discuss the impact of dead time on the detection statistics in detail. To follow the general discussion here it can be skipped though.\\
A measure for the acquisition speed is the achievable detection rate $f_{det}$. We state a linearized formula for $f_{det}$, which  illustrates the most important relations very well :
\begin{equation}
	f_{det}=\frac{1}{\frac{1}{\eta\cdot\mu\cdot\Gamma}+\tau}
	\label{fdet}
\end{equation}
where $\eta$ is the detection efficiency$, \mu$ is the incident photon flux [photons/sec], $\Gamma=f_{gate}\cdot\Delta t_{gate}$ is the detection duty cycle and $\tau$ is the dead time.\\\textit{a) high flux :} If the photon flux is large, the detection rate is limited by the dead time, i.e. $f_{det,max}=\frac{1}{\tau}$. In order to increase the detection rate, the dead time needs to be decreased. In Sect.\ref{Afterpulsing} we have already started to discuss the possibilities of afterpulse mitigation. The quenching technique, that yields to date the lowest afterpulsing, is \textit{rapid gating} \cite{Shields}\cite{Inoue}\cite{Jun2}. Dead times of the order of 10 ns can be considered realistic. This is approximately a factor 1000 better than the best actively quenched circuits can deliver. To maintain a reasonable duty cycle, gating frequencies of the order of 1 GHz are used ($\Delta t_{pulse}\approx 200$ ps).\\
\textit{b) low flux :} If the flux is small, the dead time is insignificant and $f_{det}=\eta\cdot\mu\cdot\Gamma$. Here the most important parameter is the duty cycle $\Gamma$ which should be preferably high. Increasing the duty cycle will finally lead to a situation that is called \textit{free running mode} \cite{Thew}, Fig.\ref{GSchemes}. The overbias is applied until a signal photon or a noise effect initiates an avalanche. The photon flux $\mu$ below which the \textit{free running mode} yields higher detection rates is roughly $\frac{1}{\eta\cdot\tau}$, see also App.\ref{APPG}. We see that in this case improved afterpulsing and thus smaller dead times, would extend the application range to higher photon fluxes. To date the best low afterpulsing solution for the \textit{free running mode} is the earlier discussed integrated active quenching approach \cite{Jun}. Fig.\ref{Regions} illustrates our discussion.\\
At this point we want to demonstrate, that using the \textit{free running mode} in its application regime (low flux), the acquisition speed of the conventional OTDR can be considerably outperformed.\\
Example : We want to scan a 10 km interval of a 200 km fiber ($\Rightarrow f_{pulse}=500$ Hz). We assume $\tau = 10\mu s$, $\eta=0.1$ $\Rightarrow$ the maximal flux is $\mu=10^6$ photons/s, which corresponds to a power of $P_{start}=-99$ dBm\footnote{It depends on the laser power, pulse width and the fiber link quality, to which distance this corresponds. Choosing $P_{peak}=400$ mW and assuming that this is also the effective power that reaches the fiber (no internal loss), $\Delta t_{pulse}= 100$ ns and a regular fiber behavior (loss = 0.2 dB/km), then -99 dBm correspond to backscatter power coming from a distance of 158 km.}.
At the end of the 10 km interval the power drops to about $P_{stop}=-103$ dBm  (assuming regular fiber behavior, attenuation =0.2 dB/km). From $P_{stop}$ one can infer $\hat{p}_{sig}$ (Eq.\ref{phat}) and using $\hat{p}_{dc}$= 2000 $\mbox{s}^{-1}$ we obtain $NEP_{norm}^{(pc)}= 3.6\cdot10^{-16} ~\mbox{W}/\sqrt{\mbox{Hz}}$ (see Eq.\ref{NEPnorm}), which enters into Eq.\ref{tmeas} for measurement time calculation. The bandwidth $B$ of the measurement is managed by appropriate averaging of adjacent points after the full data is acquired. Here we suppose a bandwidth $B$ of 10 MHz (averaging on 50~ns intervals) and a signal to noise ratio (SNR) of 4. All together we compute a measurement time of approximately 20 s.\\
To obtain the time of the conventional OTDR we calculate the $NEP_{norm}^{(conv)}$ and the ratio of Eq.\ref{timeratio}. In the beginning of Sec.\ref{meastime}, we stated that the $NEP_{norm,0}^{(conv)}$ was about 63-100 times larger than the $NEP_{norm,0}^{(pc)}=10^{-16}~\mbox{W}/\sqrt{\mbox{Hz}}$. The signal power is low enough to neglect the signal induced noise contribution and we obtain simply $NEP_{norm}^{(conv)}\approx 63\cdot NEP_{norm,0}^{(pc)}=6.3\cdot10^{-15}~\mbox{W}/\sqrt{\mbox{Hz}}$
$$
\Rightarrow \frac{t^{(conv)}}{t^{(pc)}}=\left(\frac{6.3\cdot10^{-15}}{3.6\cdot10^{-16}}\right)^2\approx 300
$$
yielding a total measurement time for the conventional device of about 1.5 h, assuming the same detection bandwidth $B$.\\
We stress that during this time, the conventional device obtains information not only on our 10 km measurement range, but also about the entire range before. The $\nu-$OTDR only scans the designated 10 km, but finishes doing this in around 20 s.
\section{Conclusion}
The huge advantage of Geiger-mode APDs with respect to its linear mode counterparts is the small core noise equivalent power $NEP_{norm,0}$. Our comparison of a state-of-the-art conventional OTDR (based on linear mode APD) and a photon counting OTDR (based on Geiger-mode APD) reveals a difference of roughly two orders of magnitude. We demonstrate that this resource has the potential to improve the dynamic range by 10 dB as well as the 2-point resolution by a factor of 20 (in the laser peak power limited regime). The important question is, how efficient can it be used in OTDR applications (concerning the measurement time)? To sample the backscatter of a fiber we have different possibilities at hand, i.e. \textit{train of gates} (with classical gating), \textit{free running mode} or \textit{rapid gating}. For sufficiently low backscatter power (order of -100 dBm; long fibers) we show that the \textit{free running mode} is capable of efficiently using the NEP advantage. For example, measuring a 10 km interval far down the fiber, yielded a measurement time of around 20 s, while the conventional device needs to integrate for about 1.5 h to average out the noise sufficiently.\\
At higher backscatter power, i.e. closer to the beginning of the fiber, we show that \textit{rapid gating} can largely profit from its reduced afterpulsing, which makes dead times of the order of 10 ns realistic.\\
We see a combination of \textit{rapid gating} for the beginning of the fiber and \textit{free running mode} for its end part as the currently best $\nu-$OTDR solution. Alternatively one can also imagine a hybrid of conventional OTDR for high flux and photon counting for low backscatter power, including high resolution scans with fine laser pulses on short distances.\\
Concerning dead zones, the conventional OTDR is clearly superior. It is more robust to sudden changes in backscatter power, while the $\nu-$OTDR suffers from the charge persistence effect. This effect can for example be mitigated by using an additional optical shutter.
\appendices
\section{Power measurement with gated APD}
\label{APPA}
We consider coherent light, with a mean photon number per second $\mu$, incident on the diode (detection efficiency $\eta$). We apply gates of duration $\Delta t_{gate}$, which means that on average $\mu\cdot\Delta t_{gate}$ photons hit the diode during a gate. Due to the limited detection efficiency not every photon leads to a detection. If our detector would be photon number resolving, the average number of signal detections per gate would be given by $\eta\cdot\mu\cdot\Delta t_{gate}$. Since our APD does not have this ability, all cases where more than one signal detection would occur, results in only one detection output. According to Poissonian statistics, the probability of having no signal detection is given by
$e^{-\eta\cdot\mu\cdot\Delta t_{gate}}$, hence the probability of having an APD signal detection output is given by :
\begin{equation}
p_{sig,gate}=1-e^{-\eta\cdot\mu\cdot\Delta t_{gate}}.
\label{detProb}
\end{equation}
$\mu$ can be expressed by the incident optical power $P_{opt}$ as
\begin{equation}
\mu=\frac{P_{opt}}{h\nu}
\label{muPopt}
\end{equation}
Solving for $P_{opt}$ yields
\begin{equation}
P_{opt}=\frac{-h\nu}{\eta\cdot\Delta t_{gate}}\mbox{ln}(1-p_{sig,gate})
\label{Popt}
\end{equation}
Measuring the ratio of the number of detections $N_{det}$ (including signal detections $N_{sig}$ and dark counts $N_{dc}$ ) and  the number of activated gates ($N_{gate}$), yields the signal detection probability per gate
$$p_{sig,gate}=\frac{N_{det}-N_{dc}}{N_{gate}}$$
and finally the incident optical power $P_{opt}$.\\
If the signal is weak, then it is apparent that first of all the signal needs to be separated from noise by applying a sufficiently large number of gates $N_{gate}$, see also App.\ref{APPC}. Once this is achieved, one has to consider the precision or statistical error of the result, which also is a function of $N_{gate}$. Two examples : a) $N_{gates}=10,~\Delta t_{gate}=100ns,~p_{dc,gate}=2\cdot10^{-4},~p_{sig,gate}=0.2$, the probability of a dark count to appear is negligible and we obtain $2\pm \sqrt{2} $ signal counts= total counts, from which we can calculate an optical input power lying in [0.7~pW, 5.3~pW]. b) $N_{gates}=10000$ and the other parameters like before, we obtain a number of total counts of $2002\pm \sqrt{2002} $ from which $2\pm \sqrt{2} $ are dark counts and $2000\pm \sqrt{2000} $ are signal counts. From the signal counts we infer an optical input power lying within [2.8~pW, 2.9~pW].  The statistical error of the second measurement is much smaller.\\
\section{Noise equivalent power of Geiger-mode APD}
\label{APPC}
In the treatment of the APD noise we mainly consider two contributions, i.e. the shot noise of a) the signal counts and b) the dark counts (assuming that afterpulse contributions can be neglected, for example by choosing a sufficiently large dead time).\\
Let $P_{opt}$ be the incident optical power on the diode. With the energy per photon $h\nu$, the detection efficiency $\eta$ and the gate width $\Delta t_{gate}$, we infer the signal detection probability per gate (linearized version of Eq.\ref{detProb}, for sufficiently small $P_{opt}$) :
\begin{equation}
p_{sig,gate}=\eta\cdot\frac{P_{opt}\cdot\Delta t_{gate}}{h\nu}
\label{Psig}
\end{equation}
After applying $N_{gate}$ gates of the same width the mean number of signal detections $N_{sig}$ is
\begin{equation}
N_{sig}=p_{sig,gate}\cdot N_{gate}
\label{Nsig}
\end{equation}
Assuming Poissonian statistics we  calculate the fluctuation
\begin{equation}
\Delta N_{sig} = \sqrt{p_{sig,gate}\cdot N_{gate}}
\label{aux1}
\end{equation}
The same derivation holds for the dark counts : we introduce a dark count probability per gate $p_{dc,gate}$ (which will be measured directly) leading to
\begin{equation}
\Delta N_{dc} = \sqrt{p_{dc,gate}\cdot N_{gate}}
\label{aux2}
\end{equation}
thus the total noise fluctuation :
\begin{equation}
            \Delta N_{tot}=\sqrt{\Delta N_{sig}^2+\Delta N_{dc}^2}
\label{totalNoise}
\end{equation}
The noise equivalent power ($NEP$) is inferred by calculating the optical power necessary to produce $\Delta N_{tot}$ counts when applying $N_{gate}$ gates. In order to achieve this we replace $P_{opt}$ by $NEP$ in Eq.\ref{Psig} and multiply by $N_{gate}$:
\begin{equation}
            \Delta N_{tot}=\eta\cdot\frac{NEP\cdot\Delta t_{gate}}{h\nu}\cdot N_{gate}
\end{equation}
Using Eq.\ref{aux1}-\ref{totalNoise} and solving for $NEP$ yields
\begin{equation}
NEP=\frac{h\nu}{\eta}\cdot\sqrt{\frac{p_{sig,gate}+p_{dc,gate}}{N_{gate} \Delta t_{gate}^2 }}\quad\quad [\mbox{W}]
\label{NEPAnnexe}
\end{equation}
The minimal detectable power $NEP_0$ is obtained by setting the signal shot noise contribution equal to zero :
\begin{equation}
NEP_0=\frac{h\nu}{\eta}\cdot\sqrt{\frac{p_{dc,gate}}{N_{gate} \Delta t_{gate}^2 }}\quad\quad [\mbox{W}]
\label{NEP0}
\end{equation}
The existence of $N_{gate}$ in these equations represents the iteration of a measurement and is a function of time ($N_{gate}=f_{pulse}\cdot t)$). The elemental measurement time is represented by $\Delta t_{gate}$, the duration of a single gate, which can be interpreted as the detection bandwidth via $B:=\frac{1}{2\cdot\Delta t_{gate}}$. These formulas are practical when a $NEP$ for a particular measurement needs to be calculated (see also App.\ref{APPD}). In order to obtain a formula which makes it easy to compare different detectors, we normalize with respect to $N_{gate}$ (which represents nothing else than the measurement time) and bandwidth $B$ :\\\\
\begin{equation}
NEP_{norm}=\frac{h\nu}{\eta}\cdot\sqrt{2\cdot(\hat{p}_{sig}+\hat{p}_{dc})}\quad\quad [\frac{\mbox{W}}{\sqrt{\mbox{Hz}}}]
\label{NEPnorm}
\end{equation}
and \\
\begin{equation}
NEP_{norm,0}=\frac{h\nu}{\eta}\cdot\sqrt{2\cdot\hat{p}_{dc}}\quad\quad [\frac{\mbox{W}}{\sqrt{\mbox{Hz}}}]
\label{NEPnorm0}
\end{equation}
where
\begin{equation}
\hat{p}_{dc}:=\frac{p_{dc,gate}}{\Delta t_{gate}}\quad,\quad
\hat{p}_{sig}:=\frac{p_{sig,gate}}{\Delta t_{gate}}
\label{phat}
\end{equation}
are the signal and dark count probability per gate, normalized with respect to the gate width in seconds \footnote{The relation between signal count/dark count probability per gate and gate width is almost linear over a large range of practical gate widths (typically ranging from nanoseconds to microseconds).Knowing $\hat{p}_{sig}$ and $\hat{p}_{dc}$ makes it easy to calculate the signal count and dark count probability of a particular gate of widths $\Delta t_{gate}$, just by multiplying it by $\Delta t_{gate}$.}.
\section{Dynamic Range of Optical Time Domain Reflectometer}
\label{APPD}
The strongest backscatter signal is observed right after the emission at time $t_0=\frac{\Delta l_{p}}{c}$, coming from the fiber locations within the interval $[0;\frac{\Delta l_{p}}{2}]$, where $c$ is the speed of light in the fiber and $\Delta l_{p}$ is the width of the laser pulse. The corresponding backscatter power is given by \cite{Derickson}
\begin{equation}
P_{BS,0}= S\cdot P_{0,eff} \cdot e^{-2\alpha_s L}(1-e^{-\alpha_s\Delta l_{p}})
\end{equation}
where $S$ is the fibers caption ratio, $P_{0,eff}$ is the effective laser peak power corrected for internal component (e.g. circulator) and connector loss and $\alpha_s$ the scattering coefficient.
 If we assume that $\alpha_s \Delta l_{p} << 1$, which is true in standard fiber ($\alpha_s\approx0.04 \mbox{km}^{-1}$)  and $\Delta l_{p} < 2$ km ($\Delta t_{p}<10\mu s$) we can expand the exponential and get
\begin{equation}
P_{BS,0}\approx\ S\cdot P_{0,eff}\cdot\alpha_s\cdot\Delta l_{p}	
\label{RB0}
\end{equation}
The dynamic range is then given by the ratio of $P_{BS,0}$ and the minimal detectable power $NEP_0~[W]$ (Eq.\ref{NEP0}):
\begin{equation}
	\mbox{dynR} =5 ~\mbox{log}(\frac{P_{BS,0}}{NEP_0})
\label{EquDynR}
\end{equation}
where the factor 5 accounts for the roundtrip in the fiber. Finally we obtain :
\begin{equation}
	\mbox{dynR} \approx\ 5 ~\mbox{log}(\frac{S\cdot P_{0,eff}\cdot\alpha_s\cdot\Delta l_{p}}{NEP_0})
\label{EquDynRApprox}
\end{equation}
We note that $NEP_0$ like used here, includes the measurement time and decreases $\propto\sqrt{t}$ (see also Eq.\ref{NEP0} in the case of the $\nu-$OTDR, where the measurement time $t$ is represented by the number of applied gates, $N_{gate}=f_{pulse}\cdot t$).\\
The operational definition of the dynamic range of an OTDR, given for example in \cite{Derickson}, contains a measurement time of 3 minutes. It is apparent that an extended measurement time enhances the $NEP_0$ and thus the dynamic range. In general, if the measurement time is increased by a factor $d$, the standard deviation of the noise is lowered by a factor $\sqrt{d}$ and thus the $NEP_0$ by the same amount.
\section{2-point resolution advantage of $\nu$-OTDR}
\label{APPE}
We assume an $x$ dB $\nu$-OTDR advantage in dynamic range (with respect to conventional OTDR, using the same laser pulse width $\Delta l_{p}$). Now we look for a factor $\alpha$ such that $\alpha\cdot\Delta l_{p}$ yields a reduction of the $\nu$-OTDR dynamic range by $x$ dB. Using Eq.\ref{EquDynRApprox},\ref{NEP0} and $\Delta t_{gate}=\frac{\Delta l_{p}}{c}$ (adapting laser pulse width and gate width) we have to fulfill
$$5\mbox{log}((\alpha\cdot\Delta l_{p})^{\frac{3}{2}})=5\mbox{log}((\Delta l_{p})^{\frac{3}{2}})-x$$
yielding
\begin{equation}
	\alpha=10^{-\frac{2}{15}x}
	\label{2pointAdv}
\end{equation}
For example : $x=10$ dB $\rightarrow \alpha=0.046$. Thus the $\nu$-OTDR can achieve the same dynamic range with a 20 times smaller pulse width.
\section{Signal to noise ratio as function of measurement time}
\label{APPF}
We derive a formula for the SNR ratio as a function of time from the photon counting perspective. However, the final result, after linearization will not contain any photon counting specific quantity and is therefore generally applicable.\\
We define the signal to noise ratio $(SNR)$ as the ratio of signal counts $N_{sig}$ to the total fluctuation of the counts $\Delta N_{tot}$ including fluctuation of signal and dark counts (like defined in App.\ref{APPC}).
$$SNR=\frac{N_{sig}}{\Delta N_{tot}}=\frac{p_{sig,gate}\cdot N_{gate}}{\sqrt{(p_{sig,gate}+p_{dc,gate})\cdot N_{gate}}}$$
using Eq.\ref{Nsig} and \ref{totalNoise}. Now we introduce the bandwidth normalized NEP (Eq.\ref{NEPnorm}) and use $N_{gate}=f_{pulse}\cdot t$ where $t$ is the measurement time
$$\Rightarrow SNR=\frac{\sqrt{2}\cdot h \nu}{\eta}\cdot\frac{p_{sig,gate}\cdot\sqrt{f_{pulse}\cdot t}}{NEP_{norm}\cdot\sqrt{\Delta t_{gate}}}$$
then replacing $p_{sig,gate}$ using Eq.\ref{detProb} and \ref{muPopt}:
\begin{equation}
	SNR=\frac{\sqrt{2}\cdot h \nu}{\eta}\cdot\frac{(1-e^{-\frac{\eta}{h\nu}\cdot P_{opt}\cdot\Delta t_{gate}})\cdot\sqrt{f_{pulse}\cdot   		  t}}{NEP_{norm}\cdot\sqrt{\Delta t_{gate}}}
	\end{equation}
If the optical input power is sufficiently small, the signal detection probability increases linearly with the optical power and we can expand the exponential to obtain
$$SNR=\frac{P_{opt}\cdot\sqrt{f_{pulse}\cdot t}\cdot\sqrt{2\cdot\Delta t_{gate}}}{NEP_{norm}}$$
\begin{equation}
	=\frac{P_{opt}\cdot\sqrt{f_{pulse}\cdot t}}{NEP_{norm}\cdot\sqrt{B}}\quad\quad\quad
\end{equation}
where we used $B=\frac{1}{2\cdot\Delta t_{gate}}$, the detection bandwidth.\\
This final formula is independent of any photon counting quantities and does also apply for the general case, including linear APD detection. In the linear regime there is even no such severe restrictions as in photon counting mode since much higher $P_{opt}$ can be processed.\\
On the other hand, if measurement time needs to be calculated as a function of $SNR$, we obtain straight forward
\begin{equation}
	t=\frac{1}{f_{pulse}}\cdot\left(\frac{SNR\cdot NEP_{norm}\cdot\sqrt{B}}{P_{opt}}\right)^2
	\label{tmeasSNR}
\end{equation}
\section{Train of gates discussion}
\label{APPG}
\begin{figure}[h]
\includegraphics[width=\linewidth]{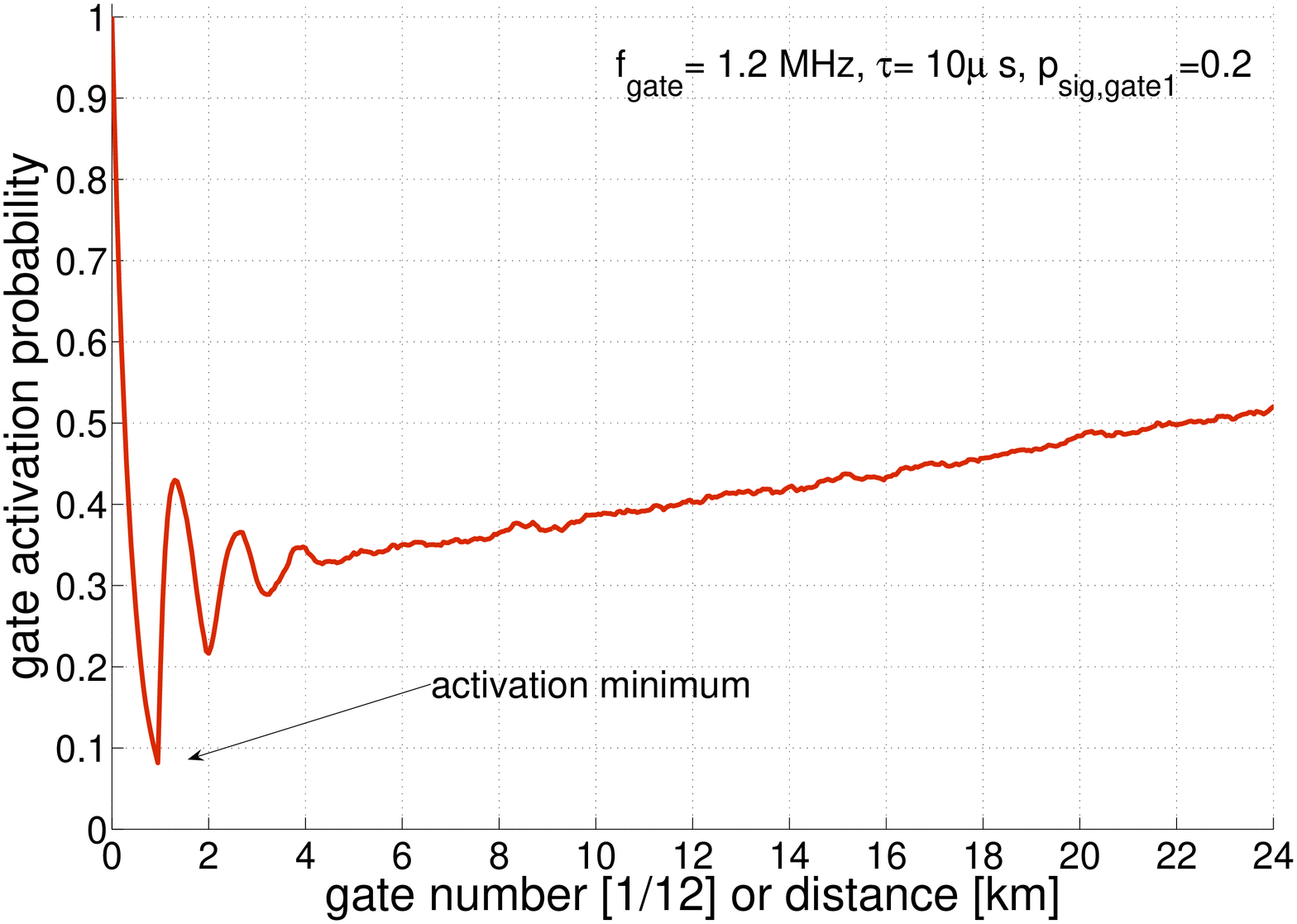}
\caption{Dead time application (here : $\tau=10 \mu s$ corresponding to 1 km in terms of sampling distance) has considerable impact on the gate activation statistics. If the gating frequency $f_{gate}$ is chosen too high, then the activation minimum approaches zero, no signal can be acquired there.}
\label{activationminimum}
\end{figure}

\begin{figure}
\centering
\includegraphics[width=\linewidth]{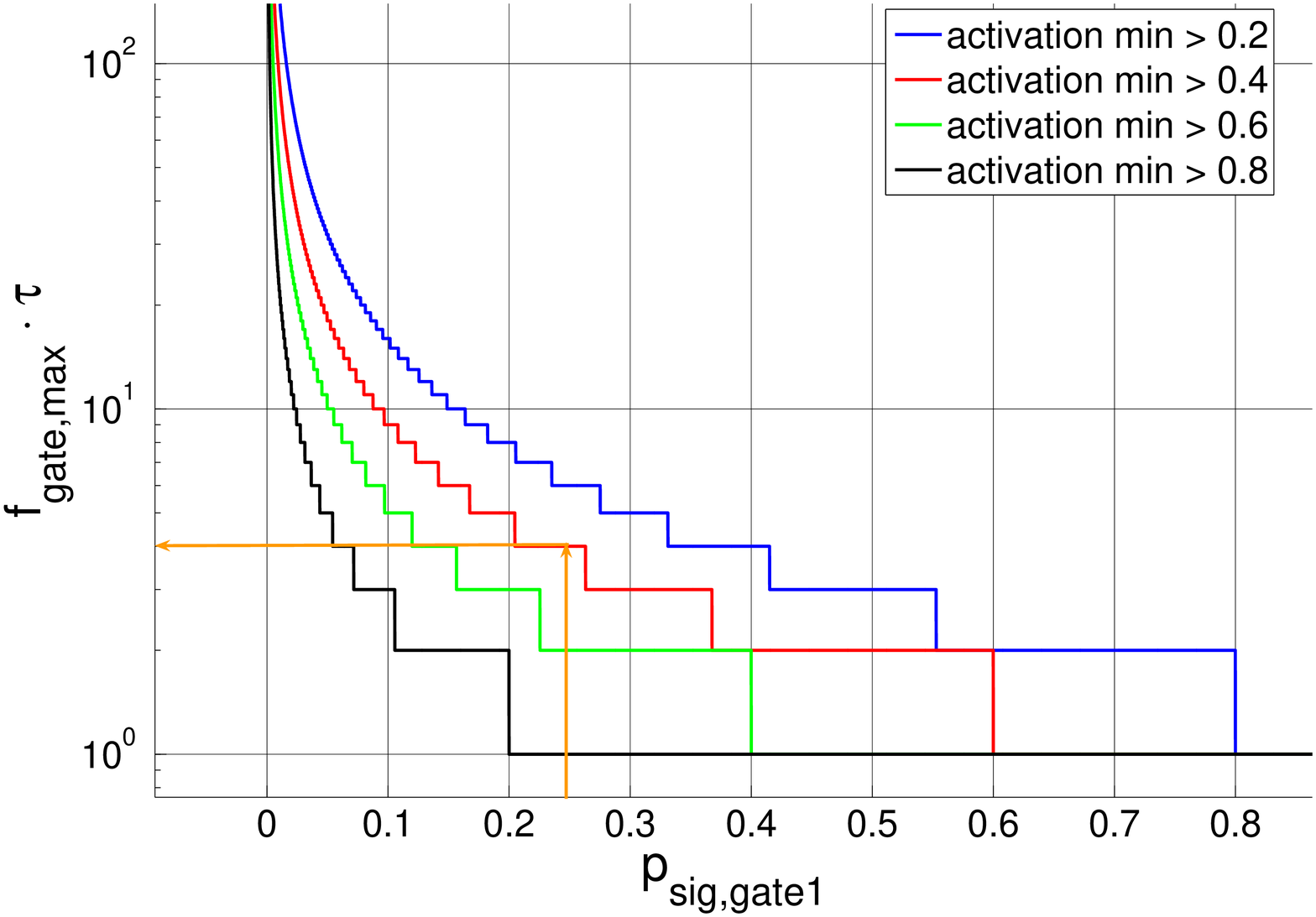} 
\caption{The maximally allowed gating frequency $f_{gate,max}$, in order to avoid activation holes, is a function of the photon detection probability in the first gate $p_{ph,gate1}$ and the designated dead time, which itself is set by the designated afterpulsing probability. Depending on the choice of the activation minimum value one can infer the product $f_{gate,max}\cdot\tau$, which also signifies the number of non activated gates after a detection. An example is given in the text.}
\label{fgatemax}
\end{figure}

The application of a dead time can have considerable impact on the gate activation statistics. Fig.\ref{activationminimum} shows what happens if the product $f_{gate}\cdot\tau$ is chosen too large.\\
If we assume, that the probability of detecting a signal photon in the first gate of the train of gates is $p_{sig,gate1}$, then gate 2 gets activated with probability $(1-p_{sig,gate1})$ (otherwise it falls into the dead time of gate 1). The probability that gate i gets activated ($i\leq f_{gate}\cdot\tau$) is then given by
\begin{equation}
	p_{act,gate~i}=(1-p_{sig,gate1})^{(i-1)}
	\label{pact}
\end{equation}
where we assume, that the signal detection probability is almost constant at the beginning of the fiber.
This expression approaches 0 when i is large. In the worst case, "activation holes" appear in a repetitive manner (see Fig.\ref{activationminimum}, in the case where the minima of the periodic structure at the beginning touch zero probability) and therefore detections around these locations are impossible or only possible with very low statistics.\\
To avoid this, we can define a criteria, which ensures that each gate has a sufficient number of activations. The first minimum plays the role of a bottleneck. When it is above some threshold (to be defined), all the other minima are as well\footnote{The reason for this is the fiber loss, which decreases the backscattered signal and therefore yields less detections and dead time applications. The dead time effect gets less severe. When the backscatter power is very low there is almost no difference to the ideal case}. The minimum depends on $p_{sig,gate1}$ and $f_{gate}\cdot\tau$.\\
Using Eq.\ref{pact} we can calculate, the maximally suitable gating frequency $f_{gate,max}$, depending on a designated threshold, see Fig.\ref{fgatemax}.\\ Example : assuming $\tau=1~ \mu s,~ p_{sig,gate1}=0.25$ and an activation minimum of 0.4. Then we infer $f_{gate,max}\cdot\tau = 4$ (see arrows in Fig.\ref{fgatemax}) and therefore $f_{gate,max}=\frac{4}{\tau}=4$ MHz.\\
If the maximally suitable gating frequency $f_{gate,max}$ is not large enough to obtain the designated sampling resolution, it is necessary to shift the start delay of the train of gates. For instance, if we want a  sampling resolution of 5 m, but $f_{gate,max}=4$ MHz, yielding only 25 m, we need to delay the train (with respect to the laser pulse departure) four times by 50 ns. This of course increases the total measurement time by a factor 5.\\
We note that $f_{gate,max}$, is in principle bounded by $1/\Delta t_{gate}$. If the suggested $f_{gate,max}$, according to Fig.\ref{fgatemax}, is larger than $1/\Delta t_{gate}$, we are naturally led to the \textit{free running mode}, where the diode stays active until a detection is obtained \cite{Thew}, see Fig.\ref{GSchemes}.\\
This happens if
\begin{equation}
	\mu<\frac{b}{\eta\cdot\tau}
		\label{fluxequ}
\end{equation}
or equivalently
\begin{equation}
	P_{opt}<h\nu\cdot\frac{b}{\eta\cdot\tau}
\end{equation}
where $\mu$ is the photon flux (number of photons per second, cw) and $P_{opt}$ the corresponding power, $\eta$ the detector efficiency, $\tau$ the detector dead time and $b$ a constant depending on the activation minimum criteria, explicitly : for an activation minimum of $0.2 (0.4,0.6,0.8)$, one obtains $b = 1.61 (0.92,0.51,0.23)$.
Due to its superior duty cycle, the \textit{free running mode} is the ideal low power solution.\\
\section*{Acknowledgment}
The authors would like to thank EXFO for providing the FTB-7600 OTDR module and support, in particular S. Perron, J. Gagnon and G. Schinn. We also like to thank J.D. Gautier for providing technical support and Bruno Sanguinetti for thorough cross-reading. We thank Swisscom for providing access to the fiber link between Geneva and Neuchatel.


\end{document}